\documentclass{article}

\usepackage{PRIMEarxiv}

\usepackage[utf8]{inputenc} % allow utf-8 input
\usepackage[T1]{fontenc}    % use 8-bit T1 fonts
\usepackage{hyperref}       % hyperlinks
\usepackage{url}            % simple URL typesetting
\usepackage{booktabs}       % professional-quality tables
\usepackage{amsfonts}       % blackboard math symbols
\usepackage{nicefrac}       % compact symbols for 1/2, etc.
\usepackage{microtype}      % microtypography
\usepackage{lipsum}
\usepackage{fancyhdr}       % header
\usepackage{multirow, multicol}
\usepackage{tabularx}
\usepackage{makecell}
\usepackage{graphicx}       % graphics
\graphicspath{{images/}}     % organize your images and other figures under media/ folder

\title{RadIA - Radio Advertisement Detection with Intelligent Analytics}

\author{J. Álvarez,
        J. C. Armenteros,
        C. Torrón,
        M. Ortega-Martín,
        A. Ardoiz,
        O. García,
        I. Arranz,\\
        \textbf{I. Galdeano,
        I. Garrido,
        A. Alonso,
        F. Bayón,
        O. Vorontsov}
        \thanks{Authors mainly are with Dezzai, C/ Pollensa n. 6, Bd. ECU 2, 2º Floor, 28290, Las Rozas, Madrid, Spain. email: \href{mailto:jorge.alvarez@dezzai.com}{jorge.alvarez@dezzai.com}, \href{mailto:juancarlos.armenteros@dezzai.com}{juancarlos.armenteros@dezzai.com}, \href{mailto:camilo.torron@dezzai.com}{camilo.torron@dezzai.com}, \href{mailto:m.ortega@ucm.es}{m.ortega@ucm.es}, \href{mailto:alfonso.ardoiz@dezzai.com}{alfonso.ardoiz@dezzai.com}, \href{mailto:oscar.garcia@dezzai.com}{oscar.garcia@dezzai.com}, \href{mailto:ignacio.arranz@dezzai.com}{ignacio.arranz@dezzai.com}, \href{mailto:i.galdeano@dezzai.com}{i.galdeano@dezzai.com}, \href{mailto:ignacio.garrido@dezzai.com}{ignacio.garrido@dezzai.com}, \href{mailto:a.alonso@dezzai.com}{a.alonso@dezzai.com}, \href{mailto:f.bayon@dezzai.com}{f.bayon@dezzai.com}, \href{mailto:o.vorontsov@dezzai.com}{o.vorontsov@dezzai.com}.
        }}

\begin{document}
\maketitle

\begin{abstract}
Radio advertising remains an integral part of modern marketing strategies, with its appeal and potential for targeted reach undeniably effective. However, the dynamic nature of radio airtime and the rising trend of multiple radio spots necessitates an efficient system for monitoring advertisement broadcasts. This study investigates a novel automated radio advertisement detection technique incorporating advanced speech recognition and text classification algorithms. RadIA's approach surpasses traditional methods by eliminating the need for prior knowledge of the broadcast content. This contribution allows for detecting \textit{impromptu} and newly introduced advertisements, providing a comprehensive solution for advertisement detection in radio broadcasting. Experimental results show that the resulting model, trained on carefully segmented and tagged text data, achieves an F1-macro score of $87.76$ against a theoretical maximum of $89.33$. This paper provides insights into the choice of hyperparameters and their impact on the model's performance. 

This study demonstrates its potential to ensure compliance with advertising broadcast contracts and offer competitive surveillance. This groundbreaking research could fundamentally change how radio advertising is monitored and open new doors for marketing optimization. 
\end{abstract}

% keywords can be removed

\section{Introduction}
\label{introduction}

Advertisements, particularly in radio broadcasts, are a significant component of modern-day social and entertainment media. They continue to play a pivotal role in marketing strategies, serving as their primary revenue source. Despite this vital role, they can often be disruptive and irritating to listeners, typically broadcast in clusters of consecutive commercials rather than in isolation.  

As the world enters an increasingly digitized era, the challenges faced by industries evolve concurrently. A notable sphere grappling with this change is the broadcasting industry, specifically in radio or audio advertisement detection. The task presents significant complexity. A proper automatic detection system contends with a sheer diversity of broadcasting styles. These range from talk radio shows to music transmissions, each offering unique scenarios the system must tackle. The tonal and thematic harmony between advertisements and the programs they punctuate further complicates this task. The challenge escalates when identifying spontaneous anchorperson marketing or novel advertisements, which dodge detection by traditional subsequence audio matching techniques due to their unpredictable nature.  
  
To monitor the presence and frequency of these ads, companies traditionally have used watermarking, a method that embeds a unique code within the advertisement content. However, this approach comes with its drawbacks. It tends to be costly and limits companies to tracking only their advertisements, offering no insight into competitors' activities.  
  
Consequently, the industry has developed techniques like audio motif discovery to detect repeating elements in audio streams. These systems primarily operate based on audio fingerprinting, maintaining a reference database filled with identifiable audio items. However, this solution presents its own set of challenges. There is a need for robust fingerprints that can withstand distortions and efficient matching methods to meet real-time requirements. 
  
Advertisement detection algorithms are broadly classified into two primary categories. The first type leverages explicit prior knowledge of an available set of advertisements and identifies them using fingerprinting methods. Meanwhile, the second type relies on heuristics as indicators of advertising content. The industry continues to innovate and evolve these approaches to meet the evolving needs of advertising monitoring.  
 
Despite the relevance of this issue, the existing literature needs to have equated methodology and standard performance metrics with evaluating detection success to compare approaches properly. The present paper proposes a strategy that does not rely on audio fingerprinting or spectral features to find previously known advertisements. It leverages state-of-the-art speech-to-text technology and supervised machine learning models to detect advertisements, regardless of their novelty, and without the necessity of identifying specific entities or brands, which may also evolve. This approach has been compared systematically, demonstrating its versatility and superiority.  
 
More specifically, state-of-the-art speech-to-text technologies (Whisper) are utilized. Whisper has been improved recently to achieve high accuracy even in noisy environments, capturing the linguistic content of the ads effectively. Experiments are designed with different sets of transcription hyperparameters for turning continuous audio chunks into segmented text units. For the text classification problem, advanced supervised learning algorithms are required. In particular, a customized RoBERTa model has shown superior performance in pattern recognition tasks. This robust combination facilitates the efficient detection of advertisements.  
 
The structure of the article is as follows:

\begin{enumerate}
    \item The first section outlines the study's motivation and the research context. It highlights the relevance of this work in the broadcasting industry and its economic impact in Spain.
    
    \item The second section presents a literature review of related works, with dedicated subsections examining spectral features, speech-to-text conversion, and text classification. 
    
    \item The third section elucidates the data acquisition process, shedding light on the quality and the extent of the curated dataset used for training and evaluating the model rigorously.  
    
    \item In the fourth section, the experimental method is illustrated with the design of various experiments related to transformers models using Whisper transcriptions. 
    
    \item The fifth section analyses and contrasts the results with comparable metrics in the literature, comprehensively evaluating the proposed methodology. 

    \item The sixth and final section presents this study's conclusions, highlighting the proposed design's competitive advantage over the current state-of-the-art, both for the specific task and for generalist models such as GPT-4. 
\end{enumerate}

\section{Motivation}
\label{Motivation}

The importance and increasing trend of radio advertising, coupled with the associated optimization of time slot segmentation and its impact on target audiences, underscores the significance of this problem. 

\begin{figure}[!t]
    \centering
    \includegraphics[width=0.475\textwidth]{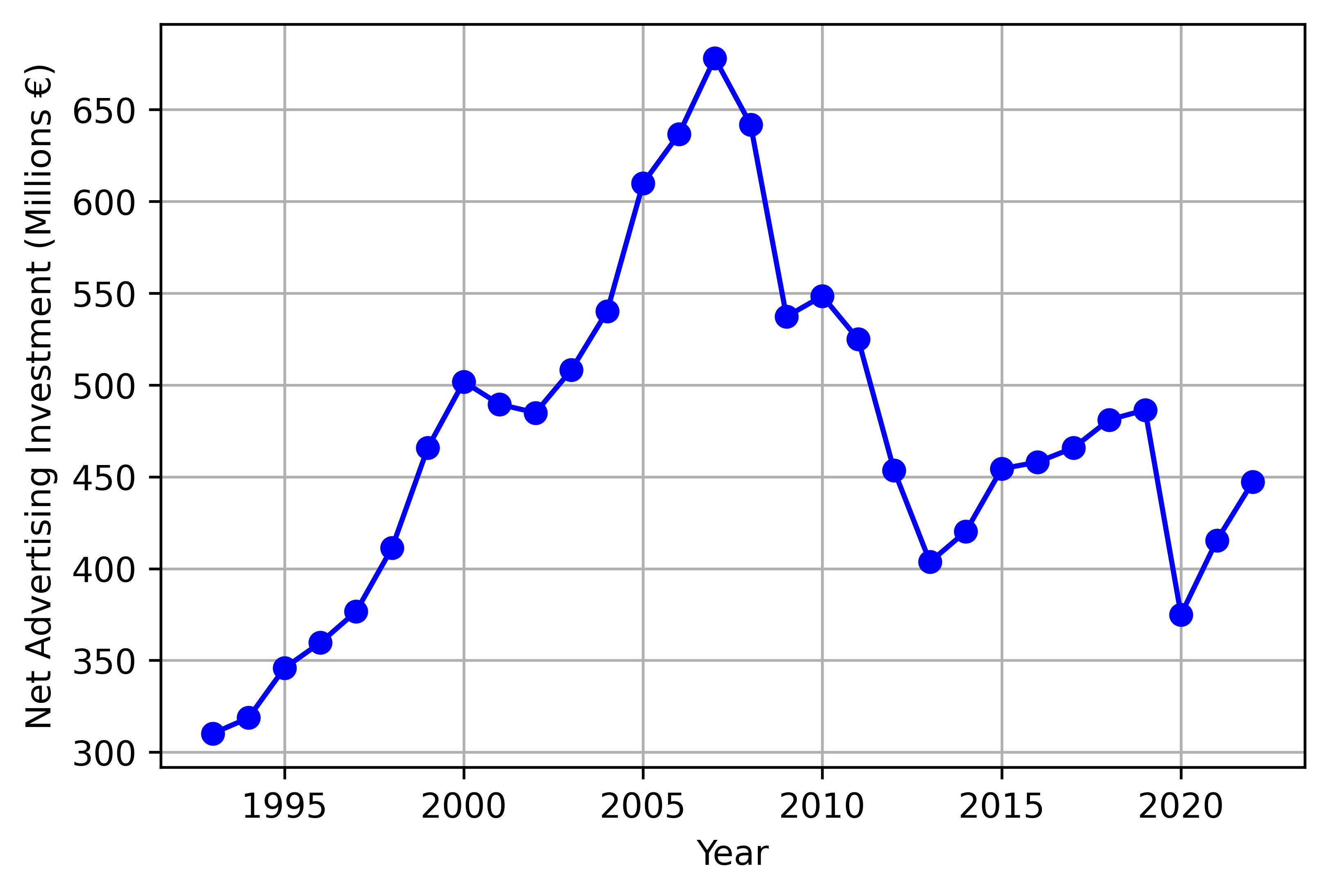}
    \caption{Net radio advertising investment evolution in Spain~\cite{infoadex}.}
    \label{fig:ad_investment}
\end{figure}

The timeline in Figure \ref{fig:ad_investment} ~\cite{infoadex} shows multiple shifts from 1993 to 2022 due to diverse factors in Spain. The economic downturn, subpar international stock market performance, the introduction of the euro, the 9/11 incidents, and a burst of the dot-com bubble caused a decline in 2000. Nevertheless, market optimism in 2003 revived investments, further boosted by significant events such as the Olympics. By 2007, the advertising sector enjoyed overall growth, driven by the rise of the Internet and changing consumer habits, causing companies to diversify their marketing strategies using a mix of traditional and modern media outlets. However, significant transformations in 2009, including mergers of large communication groups, cessation of advertising on RTVE, and analog blackout leading to audience fragmentation, resulted in a sharp decline in radio advertising. This decrease was exacerbated by the economic crisis that affected Spanish households in 2008, and the drop in investments continued until 2013. Bouncing back despite a sharp fall in 2020 due to the COVID-19 pandemic amidst economic uncertainty, the advertising sector has seen resurgence trends since then, thanks to improved economic conditions and the ongoing effectiveness of radio advertising. 
 
\begin{figure}[!t]
    \centering
    \includegraphics[width=0.475\textwidth]{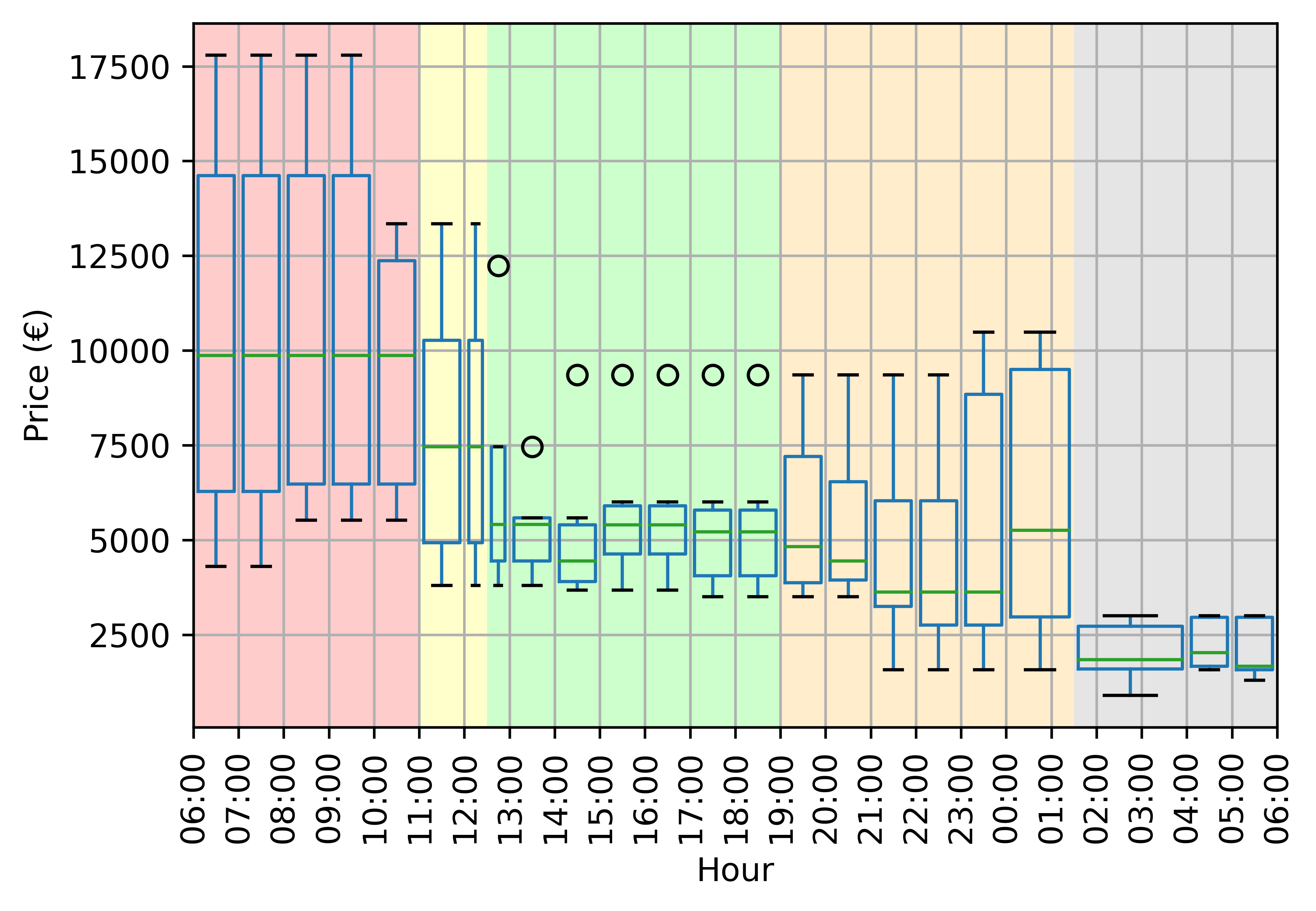}
    \caption{Radio advertising rates of a $20$-second national-wide broadcast spot during a workday in Spain updated as of 10 January 2023~\cite{oblicua}. }
    \label{fig:time-price}
\end{figure}

As reported by Oblicua~\cite{oblicua}, a renowned media agency in Spain offering advertising services across various platforms, the cost of radio advertising can vary significantly depending on the program, the time slot, and the station. Figure \ref{fig:time-price} provides a statistical summary of these costs, demonstrating the highest rates during the early morning hours, with prices gradually decreasing as the day progresses. 

\section{Related Work}
\label{Related Work}

Advertisement detection in radio streams typically focuses on identifying previously known ads. They rely on features such as the signal's spectrogram or time-varying measurements. There are two main approaches for analyzing these features~\cite{rong_audio_2016, sharma_trends_2020}. One approach involves the fingerprinting technique, while the other involves employing a machine-learning algorithm to classify the audio segments~\cite{sainath_modeling_2016, rong_audio_2016, hershey_cnn_2017}. 

\subsection{Spectral features}
\label{Spectral features}

Firstly, fingerprinting extracts features and exploits the repetition in the radio stream, such as Shazam~\cite{wang_shazam_2006} or Dejavu~\cite{dejavu}. For example, Cardinal et al.~\cite{cardinal_content-based_2010} compute the fingerprint for every frame of the audio stream. Then, try to align the fingerprint of the advertising in the audio stream for detection. 

Secondly, extracting the spectral features of the signal to feed a machine learning algorithm, such as Adblock Radio~\cite{adblock}. It uses Mel-frequency cepstral coefficients (MFCC) as input for a neural network to classify between advertising, speech, or music. Also, Koolagudi et al.~\cite{koolagudi_advertisement_2015} utilize the MFCCs as input for an ensemble of an artificial neural network and a Hidden Markov Model. 

In the study conducted by Khemiri et al.~\cite{khemiri_detection_2014}, MFCCs feed the Automatic Language Independent Speech Processing (ALISP) fingerprinting method. Subsequently, a Basic Local Alignment Search Tool (BLAST) based search is used to find repetitive audio segments in radio streams. On the other hand, Amrane et al.~\cite{amrane_deep_2022} proposed a hybrid model for advertisement detection in broadcast TV and radio content, which combines different deep learning models that take as input a combination of MFCCs and other signal features such as Root Mean Square or Tonal Centroid. Besides, the semi-supervised learning proposed by Cances et al.~\cite{cances_comparison_2022} explores the possibility of reducing the amount of labeled data to train the models. 
 
\subsection{Speech-to-text}
\label{Speech-to-text}

Rather than addressing the detection problem through signal processing techniques, an alternative approach involves leveraging advanced text-based methods for classification. One such approach is to detect advertisements directly within the textual content. For radio streams, this entails employing a Speech-to-Text or Automatic Speech Recognition (ASR) transcription module to convert the audio into text. This text-based representation can then be analyzed to identify advertising. 
 
Baevski et al. introduced Wav2Vec~\cite{baevski_wav2vec_2020}, a framework that leverages a combination of transformers and Convolutional Neural Networks (CNN) for self-supervised learning of speech representations. Learning directly from raw audio data without human annotations proves particularly advantageous in scenarios with limited labeled data. 
  
Later, Malik et al.~\cite{malik_automatic_2021} investigated various alternatives for constructing an Automatic Speech Recognition (ASR) system. Their findings indicated that the most widely used approach was combining Mel Frequency Cepstral Coefficients (MFCC) and Hidden Markov Models (HMM). Additionally, they observed that CNN and Recurrent Neural Networks (RNN) exhibited superior performance compared to other strategies in ASR tasks. 
 
Another viable option is to train a deep neural network-based speech recognition model using a dedicated speech recognition dataset, as demonstrated by Chan et al. in their work on SpeechStew~\cite{chan_speechstew_2021}. While this technique excels in generating high-quality speech representations, it may need help with generalization to datasets beyond those on which it was trained. 
 
Recently, Radford et al. introduced Whisper~\cite{radford_robust_2022}, a weakly supervised pre-training approach for speech recognition, to bridge the gap between high-quality datasets and previous unsupervised methods. Whisper was trained on a massive amount of labeled audio, spanning over $680\,000$ hours, and showcased the ability of such models to transfer effectively to different domains without requiring fine-tuning. 

\subsection{Text classification}
\label{Text classification}

Once the speech is transcribed into text, a classification system is needed to differentiate between advertisement and non-advertisement segments. Deep learning techniques have proven effective in similar tasks, as demonstrated by Guo et al.~\cite{guo_detection_2015}, who showed that a deep learning-based approach outperforms other traditional machine learning algorithms such as Support Vector Machines, Decision Trees, and Random Forests. 
 
\begin{table*}[t!]
    \caption{State-of-the-art metrics.}
    \label{tab:state-of-the-art-metrics}
    \centering
    \begin{tabularx}{\textwidth}{*{2}{l}*{4}{c}X}
        \toprule
         Authors & Technology & Accuracy & Precision & Recall & F1 & Comments \\
         \midrule
         Koolagudi et al.~\cite{koolagudi_advertisement_2015} & ANN + HMM & $87.05$ & & & & The results are for Kannada, but similar results were obtained for Hindi and English audio.  \\
         \midrule
         Khemiri et al.~\cite{khemiri_detection_2014} & ALIPS + HMM & & $99$ & $96$ & $97$ & French radios. \\
         \midrule
         \multirow[!t]{2}{*}{Amrane et al.~\cite{amrane_deep_2022}} & ANN + DTW &  & $93.35$ & $70.25$ & $80.17$ & TV and radio channel  \\
          & ANN + Autoencoder & & $92.10$ & $84.05$ & $87.66$ & streams.\\
        \bottomrule
    \end{tabularx}
\end{table*}

As shown in Table \ref{tab:state-of-the-art-metrics}, some state-of-the-art techniques can achieve great results but rely upon previously known ads. Indeed, while the approach of Khemiri et al.~\cite{khemiri_detection_2014} stands out and is particularly impressive, it is heavily dependent on finding a discernible pattern of sufficient spectral variations in the spectrogram of known advertisements. This dependency is a significant shared constraint, limiting its applicability, especially when dealing with new or concealed advertising content. The proposed approach capitalizes on both strategies: text-to-speech transcription and text classification utilizing a deep learning transformer model. Therefore, the message and its context play a crucial role.

\section{Data curation}
\label{Data curation}

Data curation in this study involved a systematic acquisition, annotation, and preparation process, ensuring the integrity and robustness of the data for subsequent analysis. 

\begin{itemize}
    \item \textbf{Data Acquisition}: Seven high-audience Spanish radio stations broadcasting in Spanish are picked for this study. The corpus compiled for this study consisted of $183$ hours and $40$ minutes of audio data, encompassing a diverse range of content from musical broadcasts to talk shows, apportioned in a $3$:$1$ ratio. The data included various recordings throughout the week and across all time slots, offering a comprehensive representation of the Spanish broadcast landscape. 
    \item \textbf{Data Annotation}: The audio data were labeled through a LabelStudio~\cite{labelstudio} session. Given the critical role of tagging in shaping the quality of results, a computational linguist is employed to secure precise and accurate annotations. The labeling mandate required audio data to be segregated into two mutually exclusive categories: advertisement (``ad") and non-advertisement (``no-ad"). Furthermore, the labeled regions were expected to cover the audio data completely. Different advertisement segments or distinctive non-advertisement collections were also segregated for further analytics.
 
    Each labeled region was assigned a confidence value to verify the fidelity of the labels in a thorough revision. This process ensured accurate attribution of uncertain regions and consistent application of the labeling standard over time. A margin of error of $100$ ms during transitions was preserved by utilizing average rounding techniques, securing complete audio coverage without overlapping to accommodate potential precision inaccuracies.

    \item \textbf{Data Preparation}: The annotation process resulted in pairs of timestamps and ``ad"/``no-ad" classifications, marking the beginning of different segments. These timestamps can be segregated into any desired temporal window resolution, offering flexibility for real-time analytical processing.
\end{itemize}

\section{Data Analytics}
\label{Data Analytics}

\begin{figure*}[!t]
    \centering
    \includegraphics[width=0.8\textwidth]{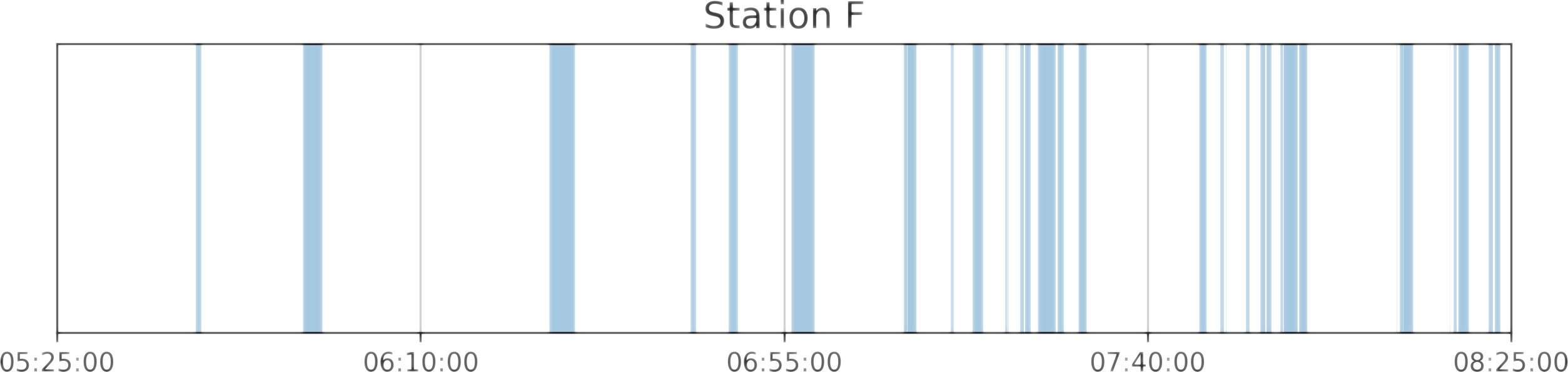}
    \caption{Example of advertisement distribution over a $3$-hour segment. Blue blocks represent time windows where advertisements are broadcasted.}
    \label{fig:example-ads}
\end{figure*}

Figure \ref{fig:example-ads} provides a comprehensive illustration of the temporal distribution of advertisements. As observed, there is no consistent pattern across time, underscoring the dynamic nature of radio advertisement strategies. This example corresponds to the peak hour early in the morning, highlighted as the most expensive time slot on the radio. 

\begin{figure}[!t]
    \centering
    \includegraphics[width=0.475\textwidth]{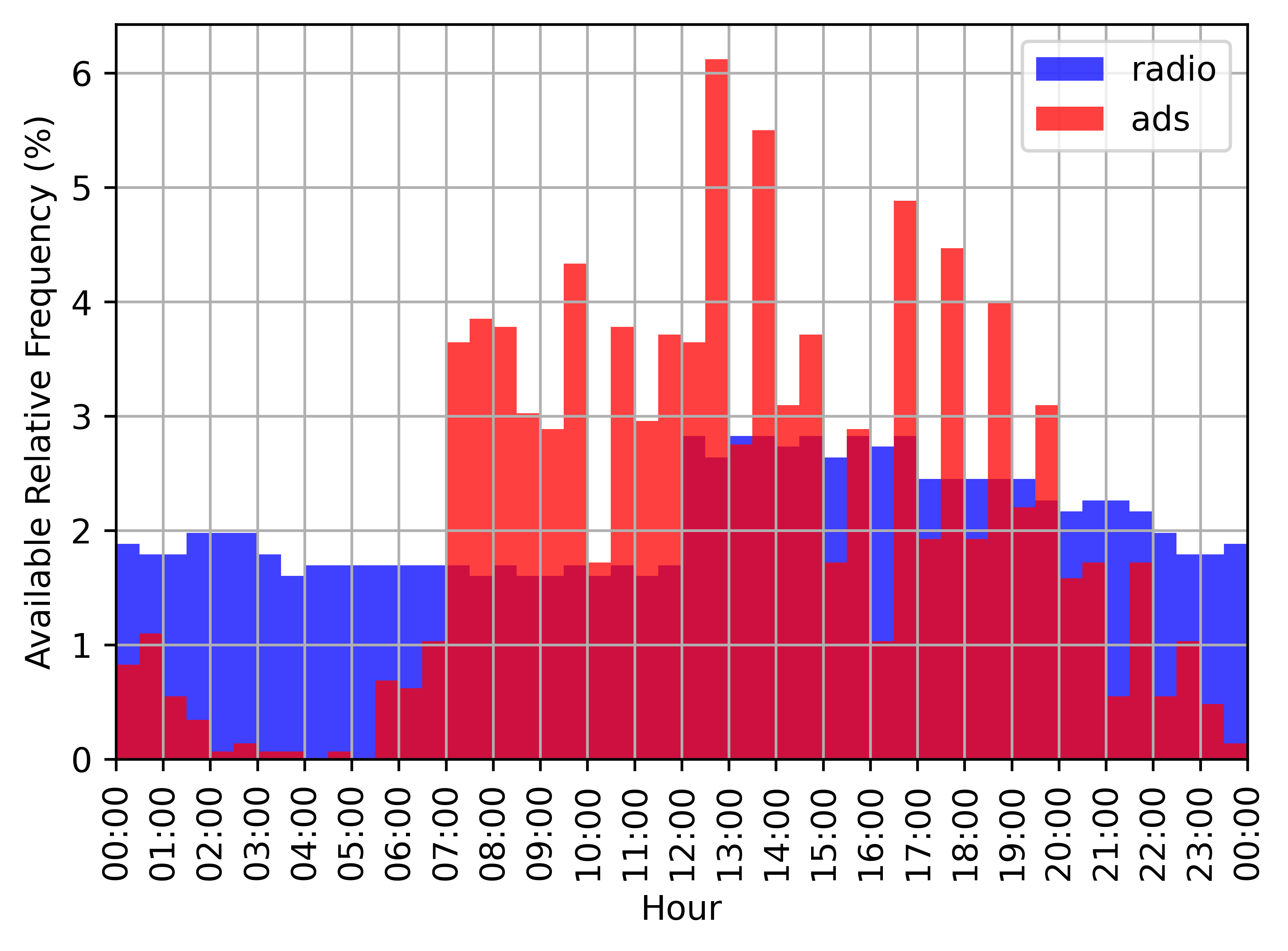}
    \caption{Time coverage of radio samples from all the stations and overall distribution of advertisements throughout the day.}
    \label{fig:available_time}
\end{figure}

Figure \ref{fig:available_time} presents the distribution of samples segregated into $30$-minute bins throughout the day. The processed radio data can be considered reasonably uniformly distributed despite a slight increase in samples in the afternoon. Nevertheless, advertisers prefer to optimize strategies between $7$ and $10$ am and during peak commuting hours. Although there is a noticeable price dip during the midday hours, possibly indicating the targeted nature of advertising, the high frequency of spots tries to maximize the return on investment. On the contrary, fewer advertisements are aired in the evening or late at night. 

\begin{figure}[!t]
    \centering
    \includegraphics[width=0.475\textwidth]{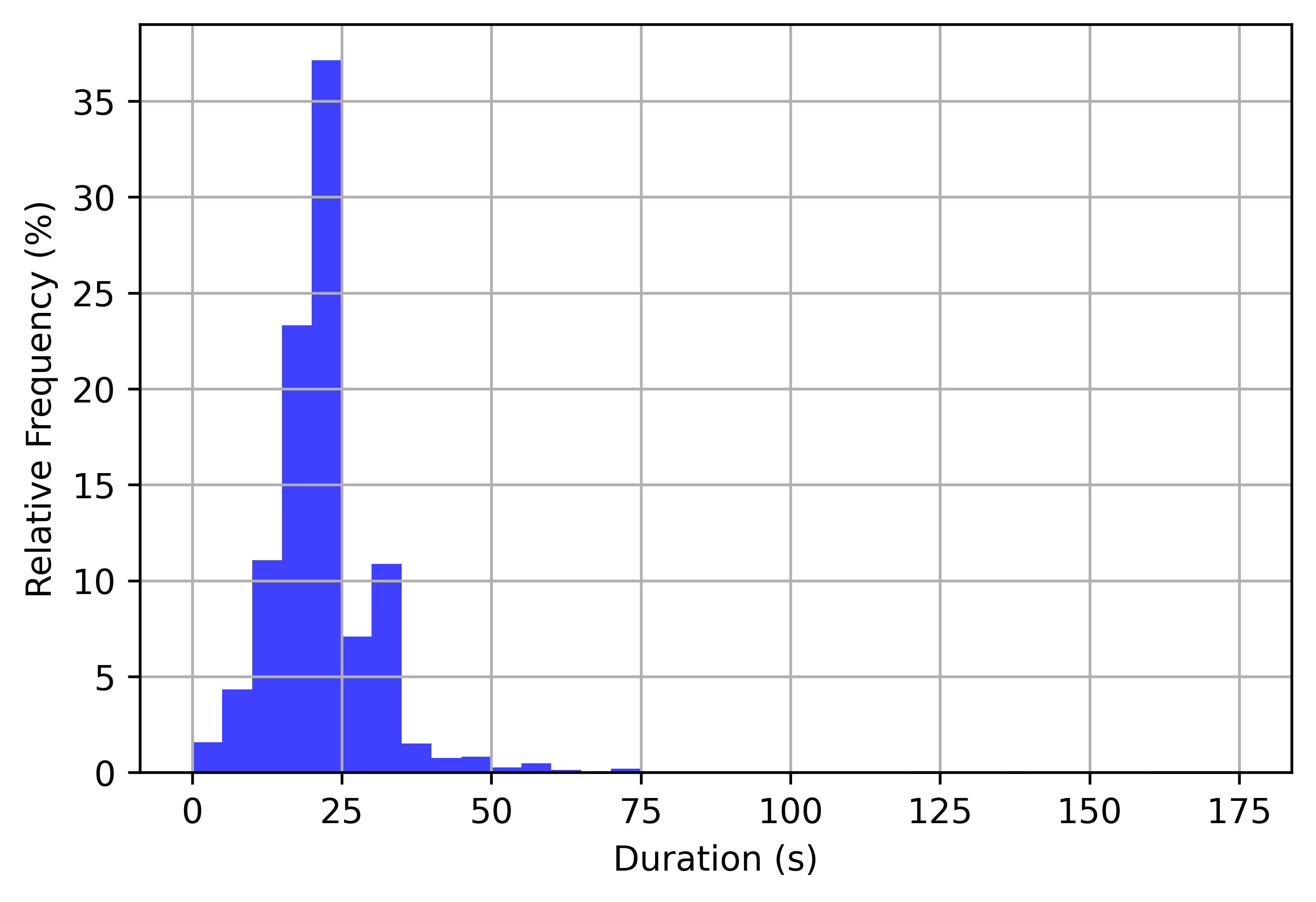}
    \caption{Time distribution of the duration of advertisements detected.}
    \label{fig:ad_distrib_time}
\end{figure}

Radio ads typically last between $20$ and $30$ seconds. However, as represented in Figure \ref{fig:ad_distrib_time}, some mentions are shorter, often seen in sports broadcasts or sponsorships, and occasionally longer, such as during an advertiser's interview or consultation moderated by the broadcaster or excerpts from other programs or songs. In any case, voice or music jingles that announce station segments or the station itself have been excluded from this analysis. 

\section{Methodology }
\label{Methodology }

The ``transcribe+classify" strategy harnesses the semantic content of the text transcribed from the audio. It should be noted that the focus on semantic content precludes the consideration of other audio attributes, such as intonation, background music, or sound effects.

This approach involves the following steps: 

\begin{enumerate}
\item \textbf{Audio-to-Text Transcription}: Transformation from the audio into text using OpenAI's Whisper. The model size can be: ``small", medium, or large, depending on the number of parameters of the architecture and its standard nomenclature. This transcription step necessitates the selection of a time window for capturing the semantic context without losing flexibility. Also, two segmentation techniques are employed for this purpose, labeled as ``non-exact" and ``exact", respectively:
    \begin{enumerate}
        \item \textbf{Non-Exact Segmentation}: Transcribe the audio in $10$-minute segments and cut the text into smaller segments using the Whisper-generated timestamps. 

        \item \textbf{Exact Segmentation}: Segment the $10$-minute audio parts into chunks of n seconds, then transcribe each chunk. $10$-minute duration of the audio is chosen given that it facilitates the work of the taggers, and longer lengths could damage the quality of the audio transcription. 
    \end{enumerate}

\item \textbf{Text Tagging}: Post-transcription, the text classification of the segments as either ``ad" or ``no-ad" for a supervised learning problem. The window tag is determined by which type of content occupies the majority of the time within that window. Consecutive windows do not overlap and leave no remainder whatsoever. Thus, the size of the window will determine the ability to discriminate advertisements in a chunk of audio.

    \item \textbf{Dataset Generation}: The advertisement detection scheme is non-sequential, i.e., each window of transcribed text is assessed independently without any continuity requirement. However, the selected test dataset complies with the following rules: 

    \begin{itemize}
        \item The test set contains segments from each station's total broadcast, with each segment comprising $3$ consecutive hours of transmission.
    
        \item The continuous $3$-hour blocks exhibit a wide range of advertisement durations (from none in late-night shows up to $16$\% of the block duration), ensuring an average advertising content of around $4$\% of the broadcasting time.

        \item At least two test blocks of $3$ hours are selected for every radio-tagged day and station (which is chosen to preserve typical train and test proportions of $4$:$1$). If under $24$ hours are tagged, select just one block.

        \item Radio data can be categorized based on themes, allowing for differentiation between stations that primarily broadcast music or talk shows and avoiding bias. Further, a 3-hour block without any advertiser is considered.
    \end{itemize}

\begin{table}[!t]
    \caption{Dataset split global details.}
    \label{tab:data_split}
    \centering
    \begin{tabular}{*{4}l}
        \toprule
         Split & Theme & Total time & Ads time \\
         \midrule
         \multirow{2}{*}{Train} & Music & $3$ days $17$ h & $3$ h $5$ min $49$ s \\
         \cmidrule{2-4}
         & Talk show & $1$ day $4$ h $50$ min & $2$ h $45$ min $2$ s \\
         \midrule
         \multirow{2}{*}{Validation} & Music & $15$ h $40$ min & $32$ min $48$ s \\
         \cmidrule{2-4}
         & Talk show & $5$ h $10$ min & $29$ min $7$ s \\
         \midrule
         \multirow{2}{*}{Test} & Music & $33$ h & $1$ h $33$ min $35$ s \\
         \cmidrule{2-4}
         & Talk show & $12$ h & $49$ min $39$ s \\
         \bottomrule
    \end{tabular}
\end{table}

\begin{table}[!t]
    \caption{Anonymized test set minor details.}
    \label{tab:test-set}
    \centering
    \begin{tabular}{*{5}l}
        \toprule
         Radio & Theme & Date & Time slot & Ads time \\
         \midrule
         Station A & Music & 12/08/2022 & 09:05-12:05 & $15$ min $48$ s \\
         \midrule
         Station B & Music & 31/05/2022 & 22:55-01:55 & $44$ s \\
         \midrule
         Station C & Music & 02/08/2022 & 00:43-03:43 & $0$ s \\
         \midrule
         Station D & Talk-show & 31/05/2022 & 14:15-17:15 & $6$ min $11$ s \\
         \midrule
         Station E & Talk-show & 31/05/2022 & 20:55-23:55 & $4$ min $38$ s \\
         \midrule
         Station F & Talk-show & 31/05/2022 & 03:55-06:55 & $8$ min $25$ s \\
         \midrule
         Station B & Music & 01/06/2022 & 12:55-15:55 & $7$ min $33$ s \\
         \midrule
         Station B & Music & 31/05/2022 & 07:15-10:15 & $12$ min $39$ s \\
         \midrule
         Station G & Music & 31/05/2022 & 21:15-00:15 & $2$ min $5$ s \\
         \midrule
         Station A & Music & 12/08/2022 & 04:45-07:45 & $6$ min $46$ s \\
         \midrule
         Station C & Music & 03/08/2022 & 06:42-09:42 & $13$ min $5$ s \\
         \midrule
         Station G & Music & 31/05/2022 & 15:25-18:25 & $12$ min $5$ s \\
         \midrule
         Station D & Talk-show & 31/05/2022 & 05:25-08:25 & $30$ min $25$ s \\
         \midrule
         Station G & Music & 01/06/2022 & 12:55-15:55 & $15$ min $29$ s \\
         \midrule
         Station G & Music & 02/06/2022 & 14:45-17:45 & $7$ min $21$ s \\
         \bottomrule
    \end{tabular}
\end{table}

    Table \ref{tab:data_split} shows the data split with the amount of audio used and their advertising time. Table \ref{tab:test-set} also displays the detailed temporal distribution of the test set, including the date and time slot of the broadcasts. Advertisements from all over the day are represented to avoid any likely bias. 

    \item \textbf{Modelling}: With train, test, and validation datasets prepared, the methodology for the text classification problem initiates. The best model is selected according to the best validation metrics during training for every transcription setup.

    \item \textbf{Evaluation}: The model's evaluation is $2$-fold: 

    \begin{itemize}
        \item \textbf{Text Classification Model Evaluation}: Target and predicted labels are compared for each instance as standard text classification. 

        \item \textbf{Approach Evaluation}: The optimal hyperparameters are sought for training and inference, separately: segmentation technique, model size, and window length. Indeed, it is feasible to train models under one transcription configuration and subsequently employ them for inference with a different transcription setup.
    \end{itemize}
\end{enumerate}

\section{Experimental setup }
\label{Experimental setup }

A transformer model appending a single output layer for binary classification (advertisement or not) is operated as the text classification model. An XLM-RoBERTa model is selected due to its state-of-the-art performance in various text classification tasks~\cite{conneau_unsupervised_2020}. Its hyperparameters include a typical configuration used in transformers, such as a learning rate of $5\times10^{-5}$, $3$ epochs, a warmup ratio of $0.1$, a weight decay of $0.01$, and an Adam epsilon of $1\times10^{-8}$~\cite{devlin_bert_2019}. A model is trained to correspond to each combination of transcription hyperparameters, accounting for $18$ models. 

This section delineates the experimental framework utilized to examine the varying parameters for audio transcription during the modeling step of the Methodology. As previously outlined, the evaluation comprises the text classification model and the assessment of the entire approach. The F1-macro score addresses the data imbalance of the broadcast duration in the classification.  

\subsection{Hyperparameter tuning}
\label{Hyperparameter tuning}

This section of the study delves into the optimization of relevant parameters of the approach:

\begin{itemize}
\item \textbf{Window Length}: Audio data was transcribed in windows of $10$, $20$, and $40$ seconds to ascertain the optimal audio duration for effective advertising classification. As previously seen, these values refer to convenient choices such as the round values for the median and areas of less than $5$\% and less than $95$\% exhibited by the advertisements in the dataset. These smaller units result in segmentation inaccuracies, as segments might include variable proportions of ``ads" and ``non-ads". The theoretical maximum F1-score gauges the best possible performance under perfect conditions within each segment. The overall effectiveness of the methodology is assessed through the actual F1-macro score, which reflects how well the model distinguishes between ads and non-ads in continuous transcriptions.

\item \textbf{Segmentation Technique}: ``Exact" and ``non-exact" segmentation techniques are experimented with for each window size. While non-exact segmentation is swifter, it is susceptible to inaccuracies as it transcribes lengthier audio segments in one go.

\item \textbf{Model Size}: Transcription was performed using $3$ different sizes of OpenAI's Whisper model: ``large-v2", ``medium" and ``small". A trade-off between transcription quality and processing time is anticipated, with smaller models being faster but potentially less accurate.
\end{itemize}

\section{Results}
\label{Results}

The discussion now transitions from the experimental setup to the consequential findings of the hyperparameter search. The corresponding results obtained from the experimentations are presented herein.

\subsection{Hyperparameter tuning}
\label{Hyperparameter tuning Results}

\begin{table}[!ht]
    \caption{Regarding various configurations, this table presents the theoretical baseline values for F1-macro and the average time necessary to transcribe $1$ hour of audio. }
    \label{tab:f1-theo}
    \centering
    \begin{tabular}{*{2}l*{3}c}
        \toprule
        \makecell{Segmentation\\ technique} & \makecell{Model\\ size} & \makecell{Window\\ (seconds)} & \makecell{Maximum\\ theoretical\\ F1-macro} & \makecell{Avg transcription\\ time for\\ $1$ h of audio}\\
        \midrule
\multirow{9}{*}{``Exact"} &
        \multirow{3}{*}{Small} & $10$ & $93.88$ & $8$ min $42$ s\\
         \cmidrule{3-5}
         &  & $20$ & $92.13$ & $6$ min $29$ s\\
         \cmidrule{3-5}
         &  & $40$ & $89.33$ & $6$ min $8$ s\\
         \cmidrule{2-5}
         & \multirow{3}{*}{Medium} & $10$ & $93.88$ & $14$ min $14$ s\\
         \cmidrule{3-5}
         &  & $20$ & $92.13$ & $10$ min $14$ s\\
         \cmidrule{3-5}
         &  & $40$ & $89.33$ & $9$ min $42$ s\\
         \cmidrule{2-5}
         & \multirow{3}{*}{Large} & $10$ & $93.88$ & $17$ min $24$ s\\
         \cmidrule{3-5}
         &  & $20$ & $92.13$ & $12$ min $12$ s\\
         \cmidrule{3-5}
         &  & $40$ & $89.33$ & $12$ min $21$ s\\
        \midrule
\multirow{9}{*}{``Non-exact"} & \multirow{3}{*}{Small} & $10$ & $93.14$ & $3$ min $53$ s\\
         \cmidrule{3-5}
         &  & $20$ & $91.58$ & $3$ min $59$ s\\
         \cmidrule{3-5}
         &  & $40$ & $88.45$ & $4$ min $20$ s\\
         \cmidrule{2-5}
         & \multirow{3}{*}{Medium} & $10$ & $93.06$ & $6$ min $51$ s\\
         \cmidrule{3-5}
         &  & $20$ & $91.35$ & $6$ min $35$ s\\
         \cmidrule{3-5}
         &  & $40$ & $88.97$ & $6$ min $51$ s\\
         \cmidrule{2-5}
         & \multirow{3}{*}{Large} & $10$ & $93.16$ & $10$ min $30$ s\\
         \cmidrule{3-5}
         &  & $20$ & $91.05$ & $10$ min $18$ s\\
         \cmidrule{3-5}
         &  & $40$ & $88.37$ & $10$ min $1$ s\\
    \bottomrule
    \end{tabular}
\end{table}

As mentioned before, the approach presented here will have an inevitable error due to the tagging of the audio-to-text windows and the binary nature of the text classification model's output. The experiments reveal that the accuracy of identifying advertisements within audio streams is influenced by window length, segmentation technique, and the model size employed for transcription. Table \ref{tab:f1-theo} illustrates the trade-off between average transcription times on an Nvidia Tesla V100 with $32$GB of VRAM and the theoretical F1-macro ceiling for every combination of the parameters. Inference times would add up to around $20$ s per audio hour using $8$ Inter Core i7 7th Gen cores.

It is imperative to emphasize that selecting the optimal model entails a multi-dimensional optimization problem encompassing six hyperparameters: training window, inference window, training segmentation, inference segmentation, training model size, and inference model size. This results in a vast space of possible combinations. Suppose the goal is to identify the best combination of hyperparameters unambiguously; exhaustive search methods or meticulous scrutiny through pinpointing inspection may prove beneficial. Nevertheless, when consistent behavior patterns emerge from summary statistics, relationships between the hyperparameters can be exploited. Leveraging these patterns can expedite decision-making and offer insights, which is advantageous for understanding performance sensitivities without exhaustive fine-tuning. 

It is worth mentioning that a grid search of values is inherently subject to a broader choice of all possible parameter values, rendering the selection of the windows impractical due to the vast number of combinations. Although there could be a loss of granularity, this strategy relies on the assumption that the patterns are consistent and meaningful and that hyperparameters' complex interaction effects do not hold. The objective must be to identify the synergistic combination that yields optimal collective performance, considering the combinatorial nature of hyperparameters. 

\begin{itemize}
    \item \textbf{Window Length}: We aim to determine if providing more contextual information during training enhances text comprehension and if shorter inference windows can be used without sacrificing performance. According to Table \ref{tab:f1-theo}, smaller window lengths yield higher theoretical F1-macro scores, indicating more precise segmentation but at the expense of losing context, which can degrade text classification performance. 

    When the length of the inference window escalates, the temporal disparities between transcriptions begin to converge. This phenomenon is more accentuated under the non-exact scheme, where the distinctions in transcription time across varying inference windows are subdued. In contrast, under the exact scheme, the transcription time for a shorter window length of $10$ seconds is observed to be substantially elongated.
    
    Models trained with longer window lengths perform worse when evaluated with shorter ones. Indeed, these models depend on a broader context for classifying texts. Conversely, models trained with shorter window lengths perform better or comparably when evaluated on test data with either the same or longer window lengths. As shown in Figure \ref{fig:train-window} and Figure \ref{fig:inference-window}, the best performance is obtained when training with the shortest window, while the inference performs best with the lengthiest window. This synergy captures the best of both worlds: the model is exposed to a concise context during training, enhancing its generalization capacity. At the same time, the elongated window during inference avails richer context adequacy that aids in making more informed predictions. Respectively, the spread of values decreases as training times are reduced or inference times increase.
    
    Besides, instead of shifting the window by its full size for each transcription step, a minor window shift during the sliding window analysis could potentially foster higher asymptotic theoretical maximum F1-macro values in Table \ref{tab:f1-theo}. Thus, it would involve analyzing progressively smaller new portions of the broadcast, potentially enhancing the precision of advertisement detection for the same window length.

    \begin{figure}[!t]
        \centering
        \includegraphics[width=0.475\textwidth]{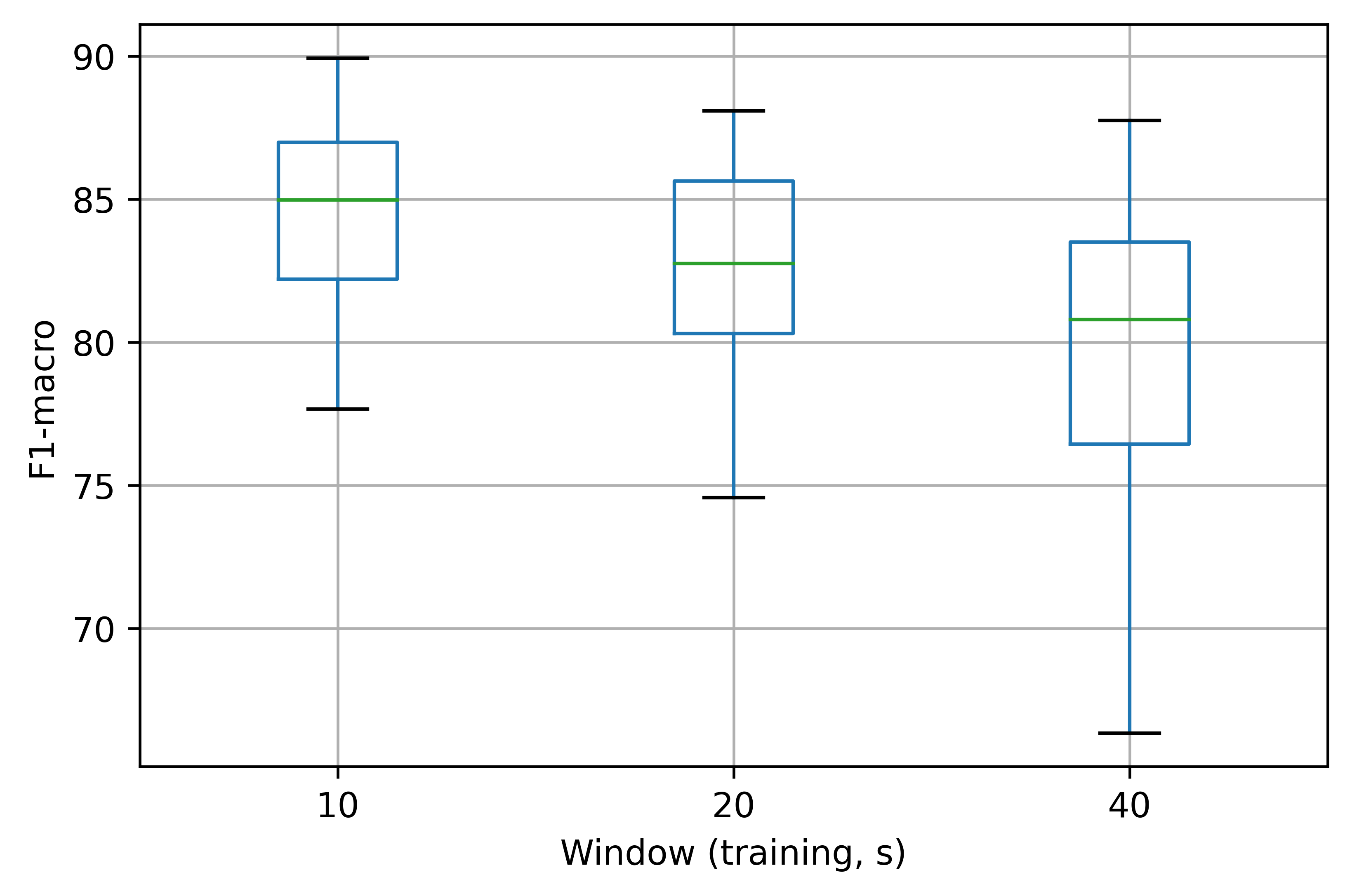}
        \caption{Boxplot of the F1-macro values for some convenient choices of training windows.}
        \label{fig:train-window}
    \end{figure}
    
    \begin{figure}[!t]
        \centering
        \includegraphics[width=0.475\textwidth]{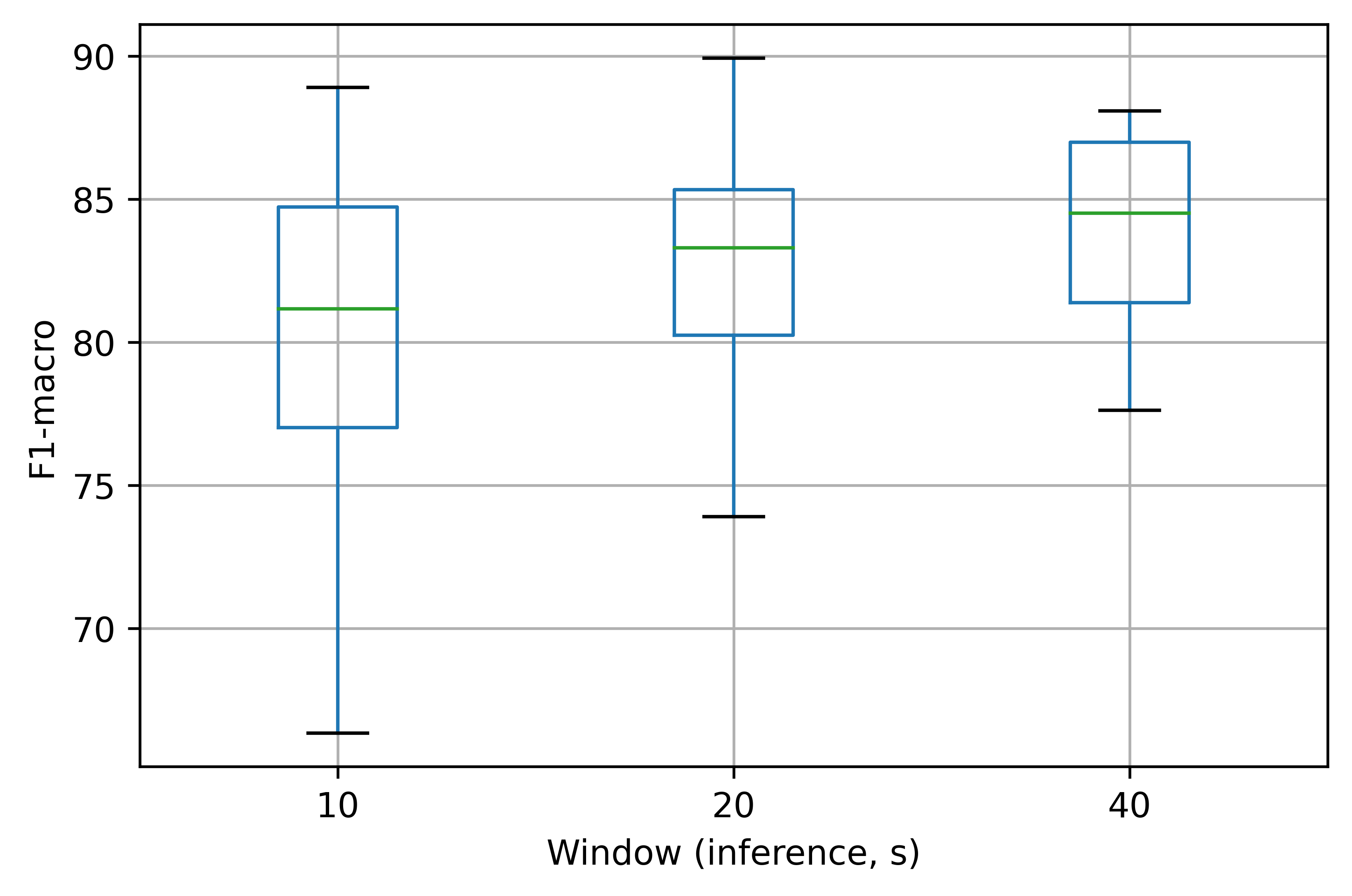}
        \caption{Boxplot of the F1-macro values for some convenient choices of inference windows.}
        \label{fig:inference-window}
    \end{figure}
    
    \begin{figure}[!t]
        \centering
        \includegraphics[width=0.475\textwidth]{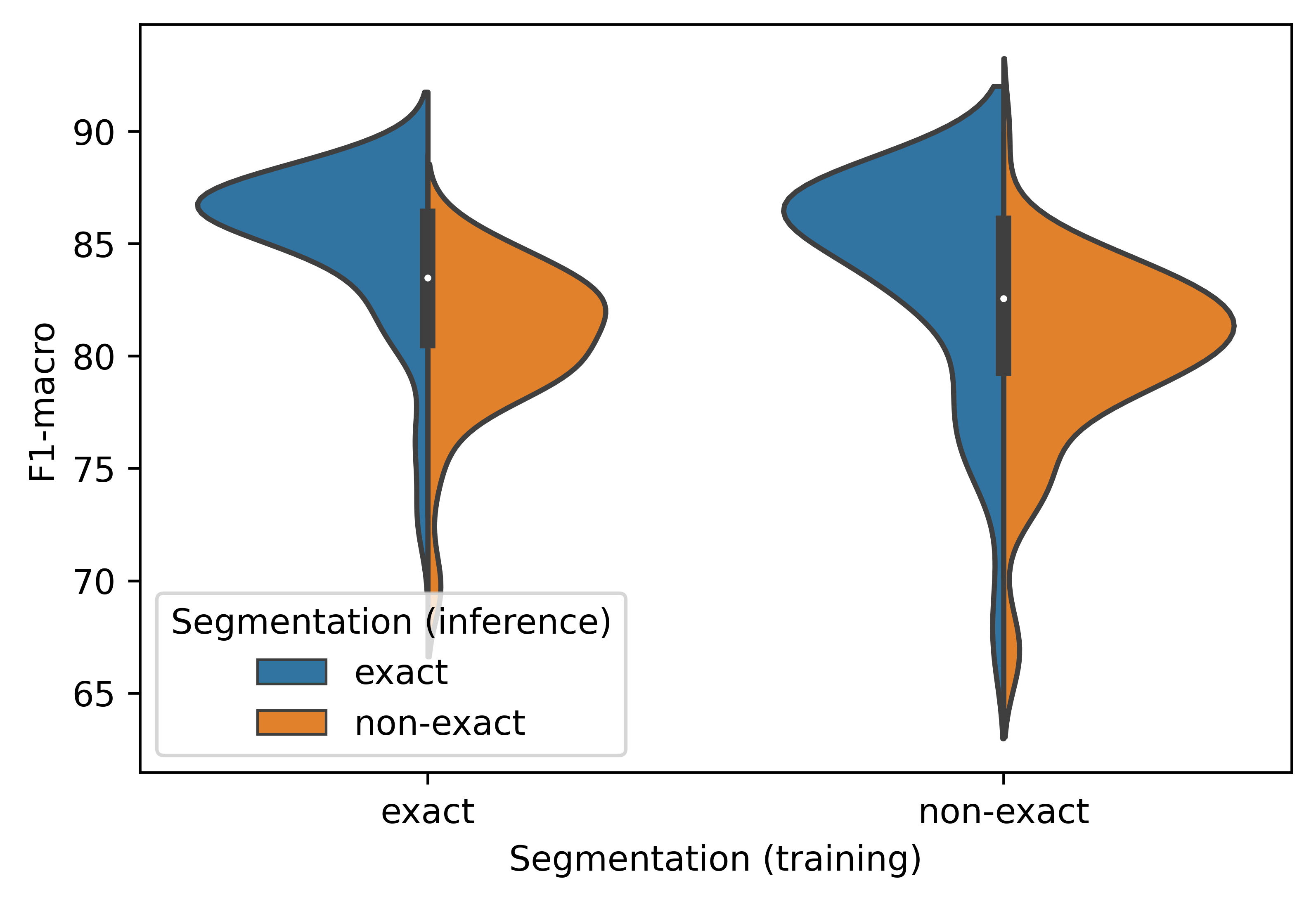}
        \caption{Violin plot of the statistical distribution of F1-macro values under different segmentation techniques.}
        \label{fig:segmentation}
    \end{figure}

\item \textbf{Segmentation Technique}: Notably, in Table \ref{tab:f1-theo}, the model size does not impact the F1-macro score when using the exact technique for the same window size. Results also indicate that the rigorous nature of the exact segmentation consistently yields better results than ``non-exact" segmentation. In Figure \ref{fig:segmentation}, a more precise segmentation instills a higher fidelity in the transcription process, thus equipping the model with a more robust understanding of the data, albeit at a higher computational cost, according to Table \ref{tab:f1-theo}. Intriguingly, the second-best performance is achieved by employing the non-exact segmentation technique during training and switching to the ``exact" segmentation technique for evaluation, trailing closely behind the scenario where exact segmentation is used for both training and inference. Despite such a slight edge, the probability density accrues predominantly at higher values with a more contained spread in the ``exact" segmentation cases than their non-exact counterparts.

It is crucial to consider that this increase in precision comes at the expense of supplementary transcriptions in the ``exact" technique. It is also worth noting that transcription errors can accumulate in longer audios in the non-exact technique, which may affect the overall accuracy of the detection process. While assigning timestamps to each word could be technically viable, achieving them precisely in practice might be challenging. Words often overlap and are spoken at varying speeds, which can complicate the exact assignment of timestamps. Post-processing techniques could be employed to adjust timestamp estimates, although this would add a layer of complexity to the process. Therefore, there is room for further optimization of the metrics obtained.

\begin{table}[!t]
    \caption{F1-macro results when the training window is $10$ s, the inference window is $40$, the training segmentation is ``exact", and the inference segmentation is ``exact" to select the best combination of training model size and inference model size}
    \label{tab:f1-size}
    \centering
    \begin{tabular}{llccc}
        \toprule
         &  & \multicolumn{3}{c}{Inference} \\
         \cmidrule{3-5}
         &  & Small & Medium & Large \\
        \midrule
        \multirow{3}{*}{Training}  & Small & $86.18$ & $86.23$ & $85.47$ \\
         \cmidrule{2-5}
         & Medium  & $87.76$ & $87.67$ & $87.52$ \\
         \cmidrule{2-5}
         & Large & $87.46$ & $86.62$ & $86.52$ \\
         \bottomrule
    \end{tabular}
\end{table}
    
\item \textbf{Model Size}: Lastly, differences among various model sizes are relatively small in Table \ref{tab:f1-size}. However, it is noteworthy that the ``medium" sized model performs the best, mainly when using the ``small" model for inference. Overall, the choice reduces to pick up the best model from Table \ref{tab:f1-size} as a subtable of Figure \ref{fig:full-table}, where the windows and segmentation techniques are fixed as previously stated.

Smaller models offer the advantage of reduced computational requirements, making them more agile and apt for scenarios where resources or time are constrained. On the other hand, larger models, while demanding greater computational power, tend to yield higher accuracy in transcriptions due to their ability to capture more complex features. As the dimensions of models escalate, they evolve into more intricate structures compared to their diminutive equivalents, thereby demanding augmented computational resources for their operation. Nonetheless, since training is an episodic undertaking, the focal point of optimization should be shifted toward streamlining the inference phase. Remarkably, the interplay between model size and other hyperparameters provides commensurate benefits to the medium (even small) sized model for tagging ``ads".
\end{itemize}

In summary, comparing the results with those of prior studies in radio advertising detection (as summarised in Table \ref{tab:state-of-the-art-metrics}), the efficacy of the methodology becomes evident. The resulting model achieves an F1-macro score of $87.76$, close to the theoretical maximum of $89.33$ that can be achieved with this time resolution, indicating high precision in advertisement detection. For practitioners aiming to harness their models' highest caliber of performance, employing a training window of $10$ seconds and an inference window of $40$ seconds and rigorously applying the exact segmentation technique in both phases is prudent. The ``medium" sized model is also advised.

\subsection{GPT-4 evaluation}

In order to validate our results, we also perform inference with the state-of-the-art Large Language Model (LLM): GPT-4. We do this over the transcriptions obtained from the 40 seconds window length, small model size, and exact segmentation technique, which, as seen before, yields the most robust results when making inferences with the model trained over the 10 seconds window length, medium model size, and exact segmentation, achieving an F1-macro score of 87.76. When evaluating this transcription configuration with \texttt{gpt-4-1106}, we get an F1-macro score of 79.12, showcasing that small models trained on a specific task can beat state-of-the-art generalist models. Furthermore, using this LLM for text classification tasks can lead to some inconveniences. During our inference process with GPT-4, we encounter some difficulties, such as the model needing to respond with the adequate output format or violating the OpenAI's responsible AI policy when inferring song lyrics, controversial radio news, or even some advertisements. Also, inference time and costs are much more significant when using GPT-4. Appendix \ref{chapter:gpt4} contains the prompt used for this evaluation.

\section{Conclusions}
\label{Conclusions}

In this research, we introduced RadIA, a novel methodology for radio advertising detection that leverages advanced natural language processing techniques. Replacing conventional methods requiring prior knowledge of broadcast content, RadIA offers a more robust and encompassing solution capable of detecting \textit{impromptu} and newly introduced advertisements. This innovation is pivotal as it transcends the limitations of conventional methods and provides a more encompassing and efficacious solution for advertising detection in the ever-evolving landscape of radio broadcasting.

A comprehensive analysis of varying transcription parameters identifies configurations delivering balanced and robust performance across different scenarios, thereby effectively assessing the trade-off between transcription quality and computational cost. Compared with the prevailing state-of-the-art techniques, RadIA demonstrates superior results, yielding an F1-macro score nearly reaching the theoretical maximum. Its several significant advantages are:

\begin{enumerate}
    \item \textbf{Discriminatory power}: The methodology does not require sufficient spectral variations to establish a differentiable pattern as it is based on textual content alone.
    \item \textbf{Scalability}: There is no need to maintain an up-to-date database of advertisements on a large scale. Besides, as the number of advertisements grows, distances become more and more costly to calculate.
    \item \textbf{Robustness}: The message prevails, and RadIA discerns advertisements covertly embedded within the presenter's comments.
\end{enumerate}

Beyond its theoretical contributions, this study underscores the practical utility of this methodology in applications where precision is paramount. It also illuminates RadIA's potential for competitive application, enabling companies to understand market activity and strategize accordingly. This dual-purpose utility, which extends to any audio broadcast (such as podcasts or web streams), offers significant added value to broadcasters and advertisers alike.

The future holds considerable anticipated implications for the broadcasting industry. RadIA has the potential to revolutionize marketing and advertising decisions, bringing a data-driven approach to strategies that have been largely qualitative until now. As we refine RadIA, we plan to explore its application to other tasks, investigate mechanisms for improving its efficiency, and experiment with different models to enhance its performance further.

\section{Acknowledgements}

This work was supported by Dezzai (\href{https://dezzai.com}{dezzai.com}). We would also like to express our sincere gratitude to Ms. Julia Sánchez Pulido (\href{mailto:julia.sanchez.pulido@gmail.com}{julia.sanchez.pulido@gmail.com}) for her invaluable contribution to this study. As a computational linguist, her expertise was instrumental in the data labeling phase of this research. Her willingness and rigorous approach to linguistic analysis greatly enriched the quality of the dataset and, by extension, the reliability of these results. The authors are deeply thankful for her commitment and assistance in this project. Her work underscores the crucial role of interdisciplinary collaboration in advancing research.

\appendix
\section{Appendix}

\subsection{General results\label{General results}}

\begin{figure*}[!t]
    \centering
    \includegraphics[width=0.85\textwidth]{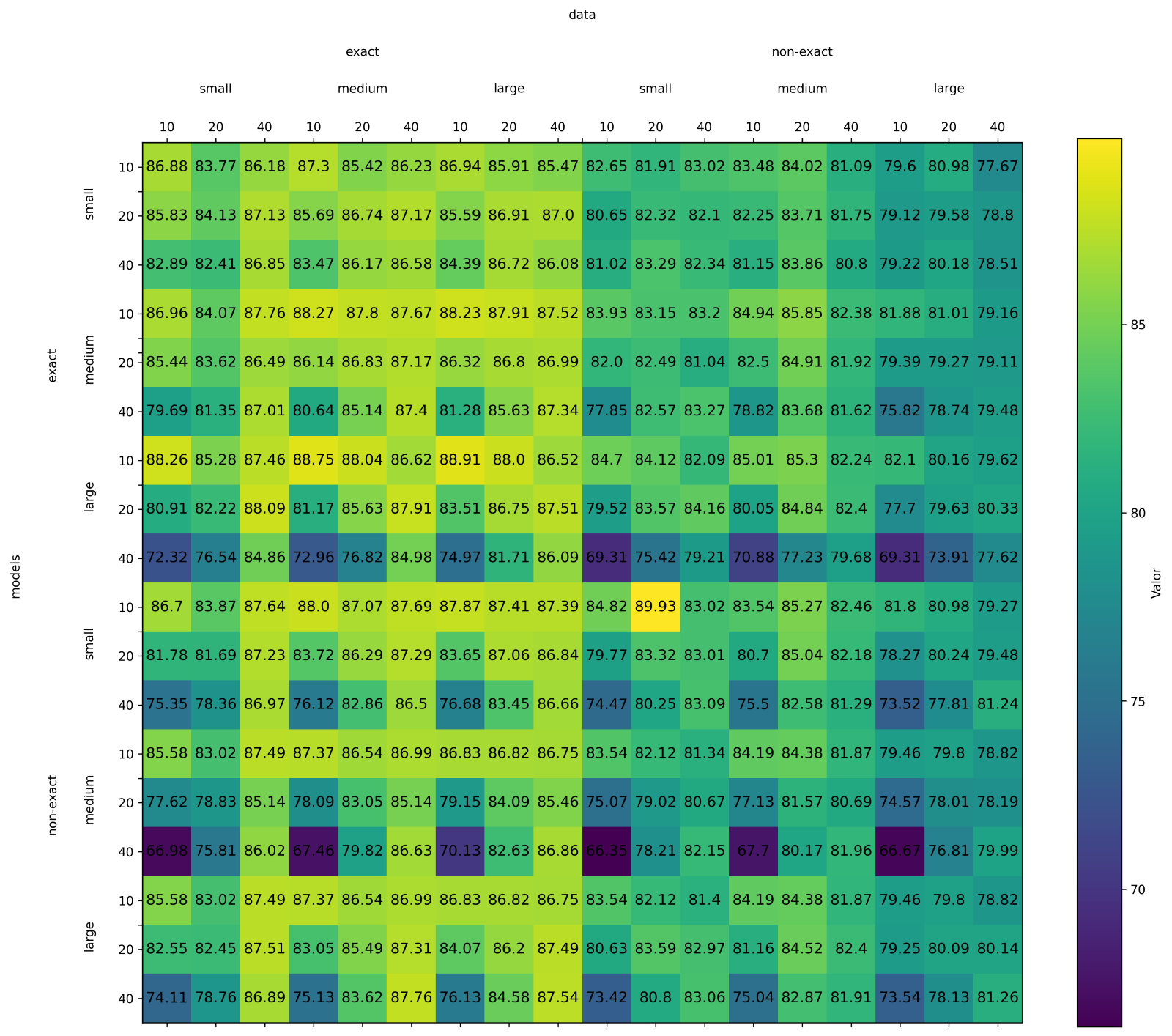}
    \caption{Heatmap of F1-macro scores illustrating the results achieved for an array of hyperparameter combinations. The color represents the magnitude of the F1-macro score, with darker shades indicating lower scores. }
    \label{fig:full-table}
\end{figure*}

Figure \ref{fig:full-table} presents a heatmap illustrating the F1-macro outcomes procured for a wide array of hyperparameter combinations. A discerning examination of Figure \ref{fig:full-table} reveals that certain amalgamations of hyperparameters yield performance metrics that surpass the previously noted F1-macro of $87.76$. However, it is prudent to exercise caution in attributing undue significance to isolated data points. A degree of inherent statistical fluctuation exists, which, while pertinent, can be judiciously overlooked in the context of the overarching behavioral trends delineated earlier. This perspective affords us the acumen to evaluate the results holistically. It precludes the proclivity to chase statistical anomalies that might not have a substantial bearing on the generalized efficacy of the model. 

We must share the original experimental results in the spirit of transparency and scholarly fidelity. This endeavor serves as an exercise in clarity and upholds the academic tenet of providing unadulterated data to the scientific community. Such openness is vital in enabling a comprehensive understanding and fostering a collaborative environment for further research and analysis. 

\subsection{GPT-4 prompt\label{chapter:gpt4}}

The prompt used in order to do the inference with GPT-4 was the following:
\\\\
\texttt{{
  "system": "Tu función es, dado un trozo de una transcripción de radio, determinar si la mayoría de ese trozo se trata de un anuncio o no."
}}
\\
\texttt{{
  "user": "Te voy a proporcionar una transcripción de un fragmento de radio de 40 segundos. Quiero que identifiques si la mayor parte de esa transcripción corresponde a un anuncio o no. En caso de que sea un anuncio de autobombo de la propia emisora, no debes considerarlo anuncio. En tu respuesta debes devolver únicamente un JSON que pueda parsear indicando si la mayor parte de esa transcripción corresponde a un anuncio o no. El formato de salida debe ser el siguiente en JSON: '\{"advertisement": "yes/no"\}'. No devuelvas un bloque de código del tipo "```json\{"advertisement":"yes"/"no"\}```". Solo el diccionario. Transcripción proporcionada:\{transcription\}'
}}

%Bibliography
\bibliographystyle{unsrt}  
\bibliography{RadIA}

\end{document}